**What we should learn from pandemic publishing**

Satyaki Sikdar[1,2,†], Sara Venturini[3,4,†], Marie-Laure Charpignon[5], Sagar Kumar[6], Francesco Rinaldi[4], Francesco Tudisco[7,8], Santo Fortunato[1,*], Maimuna S. Majumder[9,10,*]

1. Luddy School of Informatics, Computing, and Engineering, Indiana University; Bloomington, IN, USA.
2. Department of Computer Science, Loyola University Chicago; Chicago, IL, USA.
3. Senseable City Laboratory, Massachusetts Institute of Technology; Cambridge, MA, USA.
4. Department of Mathematics "Tullio Levi-Civita", University of Padova; Padova, Italy.
5. Institute for Data, Systems, and Society; Massachusetts Institute of Technology; Cambridge, MA, USA.
6. Network Science Institute, Northeastern University; Boston, MA, USA.
7. School of Mathematics, The University of Edinburgh; Edinburgh, Scotland, UK.
8. School of Mathematics, Gran Sasso Science Institute; L'Aquila, Italy.
9. Department of Pediatrics, Harvard Medical School; Boston, MA, USA.
10. Computational Health Informatics Program, Boston Children's Hospital; Boston, MA, USA.

**Corresponding authors:** Santo Fortunato santo@indiana.edu, Maimuna S. Majumder maimuna.majumder@childrens.harvard.edu

[†] These authors contributed equally to this work and are listed alphabetically by last name.

Standfirst: *Authors of COVID-19 papers produced during the pandemic were overwhelmingly not subject matter experts. Such a massive inflow of scholars from different expertise areas is both an asset and a potential problem. Domain-informed scientific collaboration is the key to preparing for future crises.*

Since the emergence of COVID-19, discussions of ongoing pandemic-related research have accounted for an unprecedented share of media coverage and debate in the public sphere[1]. The urgency of the pandemic forced researchers to operate on an accelerated timeline, as both policymakers and the public relied on the most current evidence to guide their decisions and behaviors. With high demand for rapid pandemic-related insights and lower barriers to entry via preprint servers, the volume of COVID-19 articles skyrocketed[2]. The pressing need for research triggered the participation of many researchers with expertise in the science of infectious disease outbreaks ('outbreak scientists'), who were joined by researchers from other disciplines ('bellwethers') and more junior researchers still in training ('newcomers') with the common goal of advancing the frontiers of pandemic science and informing policy decisions[3]. Please see Box 1 for details of this taxonomy.

**Collaborative efforts against COVID-19**

The scientific community's response to the COVID-19 pandemic was a highly collaborative effort[4]. This reality prompted us to investigate the allocation of human capital within and between outbreak scientists, bellwethers, and newcomers over time. We envision the ideal scenario as one where bellwethers can easily interact with outbreak scientists and engage in domain-informed collaboration. Therefore, we were particularly interested in quantifying the propensity for bellwethers to work with outbreak scientists during the COVID-19 pandemic.

The first two years of the pandemic were characterized by a rapid growth in the number of publications, followed by sustained scientific production at approximately 13,000 COVID-19-related papers per month. We used publication data from the OpenAlex database[5] to determine the composition of each paper's authoring team according to our taxonomy (i.e., outbreak scientist, bellwether, newcomer). Outbreak scientists predominantly emanated from Medicine (48%), whereas bellwethers had more diverse backgrounds like Computer Science (12%), Psychology (8%), and Business (3.4%).

**Contributions by outbreak scientists**

Between 2020 and 2022, only 7.7% of COVID-19 authors were outbreak scientists, and only 38.7% of works were contributed by teams with at least one outbreak scientist (Table 1). In the first six months, outbreak scientists accounted for 21% of all authors and contributed to 51% of papers (Fig. 1). However, their participation rapidly dwindled as bellwethers and newcomers joined the fold. Starting in January 2021, nearly two-thirds of COVID-19 papers were authored by teams in which not a single author had prior experience in outbreak science. This finding may signal the risk of misguided scientific practices during crises, as underscored by an unprecedented number of paper retractions in 2023[6]. While authors from other disciplines certainly bring fresh perspectives to the fore, domain-informed collaborations that include subject matter experts yield better situated and more creative research[7].

**Comparing COVID-19 with H1N1 and MERS**

We also examined authorship of scientific papers on two prior infectious disease crises: H1N1 influenza in 2009–2010 and Middle East Respiratory Syndrome (MERS) in 2012. In both cases, newcomers and bellwethers contributed to a substantially smaller fraction of articles than for COVID-19. This dissimilarity may partly owe to the profound, direct impact of COVID-19 on people's daily lives, in excess of that associated with H1N1 and MERS. The emergence of COVID-19 was also marked by (i) limited freedom in research topic choices because funding agencies and governments prioritized the financing of COVID-19-related research, (ii) significant barriers to the conventional execution of science (e.g., access to lab spaces and availability of supplies), and (iii) changes in publishing incentive structures and manuscript review prioritization that likely favored COVID-19 research over other topics[8].

**Fostering interdisciplinary research**

Given these data, we suggest that the COVID-19 crisis prompted many scientists to partially pivot their research activity toward topics related to the pandemic. Owing in part to disciplinary and institutional silos and in part to high demand on the time of outbreak scientists tasked to address the pandemic, bellwethers and newcomers may not have had sufficient access to subject matter experts—thus undermining opportunities for domain-informed collaboration. Therefore, analyzing the phenotypes of COVID-19 research contributors in more depth may help inform the formation and composition of interdisciplinary scientific committees and outbreak response teams in the future. To better prepare for forthcoming crises, including those beyond the realm of infectious diseases, we must make concrete investments in democratizing interdisciplinary collaboration.

We call for a concerted effort from all actors involved across various stages of the scientific ecosystem—scientists who conceive new ideas, publishers who provide platforms for knowledge dissemination, and policymakers who influence the general research agenda by controlling the allocation of resources to federal funding agencies.

*For scientists*

We encourage established researchers to connect with potential collaborators in infectious disease modeling and outbreak science, contributing their expertise to better prepare for future pandemics. Tools like [NIH Reporter](#) can help identify investigators with active grants, while platforms such as [Google Scholar](#), [ResearchGate](#), and [LinkedIn](#) can help establish new collaborations.

We also encourage researchers in training, such as doctoral and postdoctoral scholars, to leverage academic and professional mentorship opportunities at events hosted by organizations like the [Society for Epidemiologic Research](#), the [Interdisciplinary Association for Population Health Sciences](#), and [Machine Learning for Health](#). However, we recognize that financial and immigration constraints often limit participation, disproportionately affecting those from underrepresented groups.

To address these concerns, we are currently developing a free, not-for-profit, open-access platform for researchers to connect across disciplines. Our proposed 'connection recommendation' system will offer mentorship opportunities, linking trainees with mentors from diverse backgrounds and career stages. This system will also help scientists position themselves within the research collaboration ecosystem and showcase their expertise, connections, and contributions to the broader scientific network. Most importantly, by situating itself entirely online, our platform will reduce the cost of networking for underrepresented scholars—thus fostering diversity in research.

*For publishers*

In parallel, we call on publishers to introduce a mandatory *author expertise statement* in which authors would list their respective areas of expertise pertaining to the paper's subject matter—perhaps as an extension to the existing author contribution statement. Such a mandate has ample precedent, e.g., federal funding mechanisms require the inclusion of subject matter experts in investigation teams. We view this solution as complementary to the database referenced above. If journals were to require an explicit statement regarding which authors contributed which skills, then researchers would be incentivized to leverage our proposed database when expertise in a given area is lacking. Ultimately, we believe that adopting these tools and practices would stimulate domain-informed collaborations, bridge existing knowledge silos, and lead to more transparent science.

*For policymakers*

Interdisciplinary scholars are uniquely positioned to function as knowledge brokers. Unfortunately, they must often overcome challenges at the beginning of their careers due to the initially lower impact of their publications[10]. However, identifying and supporting these promising talents early on manifests in a greater return-on-investment for funders in the long term compared to their more siloed counterparts[10]. More than a decade ago, the NIH launched a visionary plan named the Common Fund to change academic culture, encourage interdisciplinary approaches, and foster team science spanning multiple biomedical and behavioral sciences. In parallel, the NSF has prioritized interdisciplinary science through solicited and unsolicited programs. The patterns of pandemic publishing indicate that these early efforts must now be expanded to stimulate, sustain, and support interdisciplinary research. This objective can be achieved by adopting long-term policy reforms and creating new research programs that foster team science across disciplines. We also call for enhanced support for scientometric research like the NSF/NIH SoS:BIO, which will help identify systemic inefficiencies and inequities and promote healthy scientific practices instead.

**Conclusion**

Amid rising concerns about reproducibility[9] and retractions[6], knowledge transfer between subject matter experts and non-experts is essential to ensure the quality and relevance of publications—particularly during crises like the COVID-19 pandemic. Especially as bellwethers foray into disciplines that are new to them, access to researchers with prior knowledge can improve their chances of making a meaningful contribution. When access to subject matter experts is limited, the quality of research may be undermined due to the authors' overreliance on incomplete domain knowledge, or the adoption of unethical scientific practices driven by pressures to publish. Such behaviors can, in turn, cause the public to cast doubt on the validity of scientific findings, possibly adding unnecessary barriers to their practical implementation and even diminishing the credibility of scientific institutions. Going forward, we hope the combination of scientist-led initiatives, technology-based solutions, editorial policies, and funding initiatives proposed here will encourage interdisciplinary research collaborations and help rebuild trust—both within the scientific community and with the public.

**Box 1**

We define three groups of authors:

- *outbreak scientists*: researchers belonging to the outbreak science community, i.e., specializing in outbreaks and infectious disease epidemiology;

- *bellwethers*: researchers from fields other than outbreak science and infectious disease epidemiology;

- *newcomers:* younger researchers still in training.

The status of researchers was ascertained based on papers they published during the pre-pandemic period (2015–2019). During the pandemic (2020–2022), the status of authors is treated as static. Specifically,

- *outbreak scientists* have authored *at least* one paper on outbreaks or infectious disease epidemiology in the pre-pandemic period;

- *bellwethers* have written at least one paper in the pre-pandemic period but *none* on outbreaks or infectious disease epidemiology;

- *newcomers* did not write any papers during the pre-pandemic period.

## Competing interests

The authors declare no competing interests.

## Acknowledgments

This research was supported in part by Lilly Endowment, Inc., through its support for the Indiana University Pervasive Technology Institute.

The authors were supported in part by the following grants. S.S.: #FA9550-19-1-0354, US Air Force Office of Scientific Research (AFOSR); S.V.: #1927425, #1927418, US National Science Foundation (NSF); S.K.: #SES2200228 (NSF); S.F.: #1927425 (NSF), #1927418 (NSF), and #FA9550-19-1-0354 (AFOSR); M.S.M.: #R35GM146974, National Institute of General Medical Sciences, National Institutes of Health. M.C. was supported by a doctoral fellowship from the Eric & Wendy Schmidt Center of the MIT-Harvard Broad Institute.

## Data and materials availability

All data, code, and materials used in the analysis are hosted on OSF.

**Figure caption**

The COVID-19 research landscape. (a) Fraction of authors in the three categories (i.e., outbreak scientists, bellwethers, newcomers) during the observation window, i.e., 2020–2022. (b) Fraction of COVID-19 papers authored by teams with a proportion of outbreak scientists (OS) ranging from 0% (No OS, in teal), to 1–50% (Minority OS, purple), to 51–99% (Majority OS, yellow), and 100% (Only OS, red). Dashed lines in the panels mark when the WHO declared COVID-19 a pandemic. For clarity, only percentages ≥ 10% are annotated in the area plots.

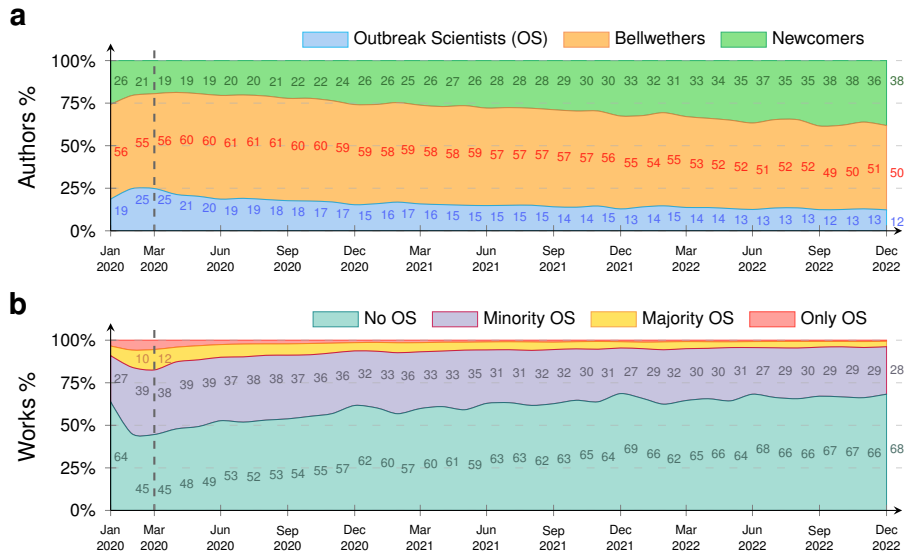

Table 1. Authorship statistics of COVID-19-related works. #Authors: the number of distinct authors by group. # Works: the number of works with at least one author from the considered group. Note that a work can count towards multiple groups (e.g., if one of the authors belongs to the group of outbreak scientists while another author is a newcomer).

|  | Outbreak Scientists | Bellwethers | Newcomers | Total |
| --- | --- | --- | --- | --- |
| #Authors | 100,736 (7.71%) | 679,424 (52.01%) | 526,070 (40.27%) | 1,306,230 |
| #Works | 175,794 (38.70%) | 408,937 (90.03%) | 301,184 (66.30%) | 454,242 |